\documentclass[slac_one]{revtex4}
\usepackage{graphicx}
\usepackage{fancyhdr}
\RequirePackage{xspace}
\usepackage{relsize}
\def\babar{\mbox{\slshape B\kern-0.1em{\smaller A}\kern-0.1em
    B\kern-0.1em{\smaller A\kern-0.2em R}}}
\mathchardef\Upsilon="7107
\def\Y#1S{\ensuremath{\Upsilon{(#1S)}}\xspace}

\def\epem{\ensuremath{e^+e^-}\xspace}
\def\Vud  {\ensuremath{|V_{ud}|}\xspace}
\def\Vus  {\ensuremath{|V_{us}|}\xspace}

\def\mtau{\ensuremath{\tau}\xspace}

\def\ms{\ensuremath{m_s}\xspace}

\def\tautau{\ensuremath{\tau^+\tau^-}\xspace}
\newcommand{\eett}   {\ensuremath{e^+e^- \to \tautau}\xspace}

\def\invfb   {\ensuremath{\mbox{\,fb}^{-1}}\xspace}

\newcommand{\gev}{\ensuremath{\mathrm{\,Ge\kern -0.1em V}}\xspace}
\newcommand{\mev}{\ensuremath{\mathrm{\,Me\kern -0.1em V}}\xspace}
\newcommand{\mevcc}{\ensuremath{{\mathrm{\,Me\kern -0.1em V\!/}c^2}}\xspace}

\def\piz   {\ensuremath{\pi^0}\xspace}

\newcommand{\tauknu}   {\ensuremath{ \tau^{-} \to K^{-} \nut}\xspace}
\newcommand{\taupinu}   {\ensuremath{ \tau^{-} \to \pi^{-} \nut}\xspace}
\newcommand{\tauenu}   {\ensuremath{ \tau^{-} \to e^{-} \nueb \nut}\xspace}
\newcommand{\taumunu}  {\ensuremath{ \tau^{-} \to \mu^{-} \numb \nut}\xspace}

\newcommand{\BFtautoknu}    {\ensuremath{\BR(\tauknu)}\xspace}                                  
\newcommand{\BFtautopinu}    {\ensuremath{\BR(\taupinu)}\xspace}
\newcommand{\BFtautoenu}    {\ensuremath{\BR(\tauenu)}\xspace}                                          
\newcommand{\BFtautomunu}    {\ensuremath{\BR(\taumunu)}\xspace}

\newcommand{\BFKmutwo}      {\ensuremath{\BR(K^- \to \mu^- \numb)}\xspace}
\newcommand{\BFpimutwo}    {\ensuremath{\BR(\pi^- \to \mu^- \numb)}\xspace}

\newcommand{\BRtautomunu}    {\ensuremath{\frac{\BR(\taumunu)}{\BR(\tauenu)} }\xspace}
\def\Kbar  {\kern 0.2em\overline{\kern -0.2em K}{}\xspace}

\def\Kz    {\ensuremath{K^0}\xspace}
\def\Kzb   {\ensuremath{\Kbar^0}\xspace}
\def\KzKzb {\ensuremath{\Kz \kern -0.16em \Kzb}\xspace}
\def\Kp    {\ensuremath{K^+}\xspace}
\def\Km    {\ensuremath{K^-}\xspace}

\def\KpKm  {\ensuremath{\Kp \kern -0.16em \Km}\xspace}
\def\KS    {\ensuremath{K^0_{\scriptscriptstyle S}}\xspace}

\def\nut        {\ensuremath{\nu_\tau}\xspace}
\def\BR         {{\ensuremath{\cal B}\xspace}}

\def\nub        {\ensuremath{\overline{\nu}}\xspace}

\def\nub        {\ensuremath{\overline{\nu}}\xspace}

\def\nueb       {\ensuremath{\nub_e}\xspace}

\def\numb       {\ensuremath{\nub_\mu}\xspace}

\def\nut        {\ensuremath{\nu_\tau}\xspace}

\newcommand{\BRknu}   {\ensuremath{ (0.03882 \pm 0.00032 (\rm{stat}) \pm 0.00056 (\rm{syst}))}\xspace}
\newcommand{\BRpinu}   {\ensuremath{(0.5945 \pm 0.0014 (\rm{stat}) \pm 0.0061 (\rm{syst}))}\xspace}

\newcommand{\BRmunu}   {\ensuremath{(0.9796 \pm 0.0016 (\rm{stat})  \pm 0.0035 (\rm{syst}))}\xspace}
\newcommand{\BRkpi}   {\ensuremath{(0.06531 \pm 0.00056 (\rm{stat}) \pm 0.00093 (\rm{syst}))}\xspace}

\newcommand{\Vudvalue} {\ensuremath{0.97408 \pm 0.00026}\xspace}

\newcommand{\gmge} {\ensuremath{|\frac{g_\mu}{g_e}|}\xspace}
\newcommand{\gmgevalue} {\ensuremath{1.0036 \pm 0.0020}\xspace}
\newcommand{\gtgm} {\ensuremath{|\frac{g_\tau}{g_\mu}|}\xspace}
\newcommand{\gtgmpvalue} {\ensuremath{0.9859 \pm 0.0057}\xspace}
\newcommand{\gtgmkvalue} {\ensuremath{0.9836 \pm 0.0087}\xspace}

\begin{document}

\title{\boldmath Lepton Universality, \Vus and search for second class current in \mtau decays} 

%

\author{Swagato Banerjee\\
 (Representing the \babar\ collaboration)}
\affiliation{University of Victoria, P.O. Box 3055, Victoria, B.C., CANADA  V8W 3P6.}

\begin{abstract}
Several hundred million $\tau$ decays have been studied with the \babar\ detector 
at the PEP-II \epem collider at the SLAC National Accelerator Laboratory.
Recent results on Charged Current Lepton Universality and two independent measurements of $|V_{us}|$
using $\tau^- \to e^-\nueb\nut$, $\mu^-\numb\nut$, $\pi^-\nut$, $K^-\nut$, and $\KS\pi^-\nut$ decays,
and a search for Second Class Current in $\tau^-\to\pi^-\omega\nut$ decays are presented,
where the charge conjugate decay modes are also implied.
\end{abstract}

\maketitle

\thispagestyle{fancy}


\section{INTRODUCTION} 
$\tau$ decays provide a clean laboratory for studying both the charged leptonic and hadronic weak currents. 
The weak interaction coupling strength between the first and second quark generations are probed with
improved precision in hadronic \mtau decays having net strangeness of unity in the final state
using \eett data collected by the \babar\ detector at the PEP II B-factory at SLAC 
at a center-of-mass energy near 10.58\gev.
The contribution due to second class current in $\tau^-\to\pi^-\omega\nut$ decays
are also probed with an order of magnitude improvement in the upper limit.

\section{\boldmath LEPTON UNIVERSALITY and $|V_{us}|$}
Tau decays into a single charged particle and neutrino(s) can be used to 
test the assumption that all the three leptons have
equal coupling strength ($g_\ell)$ to the charged gauge bosons of the electro-weak interaction, 
known as charged current lepton universality~\cite{lepuniv}. 
While a precise measurement of the ratio $\BRtautomunu$ tests $\mu-$e charged current lepton universality,
the ratio $\frac{\BFtautopinu}{\BFpimutwo}$ tests $\tau-\mu$ charged current lepton universality with light mesons.
Similarly, a precise measurement of the ratio $\frac{\BFtautoknu}{\BFKmutwo}$ tests $\tau-\mu$ universality with strange mesons.

The largest off-diagonal element, \Vus, of the Cabibbo-Kobayashi-Maskawa (CKM) quark mixing matrix~\cite{CKM}
can be measured from the ratio of strange to non-strange inclusive branching fractions of the \mtau lepton,
interpreted in the framework of the Operator Product Expansion (OPE) and finite energy sum rules (FESR)~\cite{Gamiz:2002nu}. 
The ratio $\frac{\BFtautoknu}{\BFtautopinu}$ also provides an alternate measurement of \Vus from $\tau$ decays, 
which is independent of the convergence of the OPE.

\subsection{\boldmath \mtau branching fractions}
Recently, the \babar\ and Belle experiments have published measurements of 
the following \mtau branching fractions with net strangeness of unity in the final state:
$\BR(\tau^- \to K^- \piz \nu_\tau)$~\cite{Aubert:2007jh},
$\BR(\tau^- \to \Kzb  \pi^- \nu_\tau)$~\cite{Epifanov:2007rf},
$\BR(\tau^- \to  K^-  \pi^- \pi^+ \nu_\tau)$~\cite{Aubert:2007mh}, and
$\BR(\tau^- \to  K^-  \phi \nu_\tau)$~\cite{Inami:2006vd,Aubert:2007mh}.

Here we report on the following new \babar\ measurements~\cite{Aubert:2008an}:
\begin{eqnarray}
\BFtautomunu/\BFtautoenu & = & \BRmunu, ~\label{br_tautomu_tautoe}\\
\BFtautopinu/\BFtautoenu & = & \BRpinu, ~\label{br_tautopi_tautoe}\\
\BFtautoknu/\BFtautoenu  & = & \BRknu, ~\label{br_tautok_tautoe}\\
\BR(\tau^-\to\bar{K}^0\pi^-\nut) & = & (0.840 \pm 0.004 (\rm{stat}) \pm 0.023 (\rm{syst}))\%.
\end{eqnarray}

The world average values of the strange branching fractions updated from previous estimates~\cite{Davier:2005xq} 
are listed in Table~\ref{Table:StrangeBranchingFractions},
which includes the new Belle measurements of 
$\BR(\tau^- \to  K^-  \pi^- \pi^+ \nu_\tau)$~\cite{Inami:2008kt}
and the modes containing an $\eta$ meson~\cite{Inami:2008ar}.
The \babar\ value of \BFtautoknu is obtained by multiplying Eqn.~\ref{br_tautok_tautoe}
by the world average measured value of $\BFtautoenu = (17.82\pm0.05)\%$~\cite{Yao:2006px}.
A scale factor $S = \sqrt{\chi^2/(M-1)}$ is evaluated using $M$ measurements with error less than $3\sqrt{N}\delta$,
where $\delta$ is the unscaled error on the mean of all the $N$ measurements.
If $S$ is greater than unity, the average error is scaled by $S$~\cite{Rosenfeld:1975fy}. 
The sum total is calculated using the world average measured value of \BFtautoknu,
and also the one predicted from $K_{\mu 2}$ decays assuming $\tau-\mu$ universality~\cite{Gamiz:2007qs}.

\begin{table}[!hbtp] 
\caption{The world average values of the strange branching fractions ${\cal{B}}(\tau^-\to X^-_{us}\nu_\tau)$.}
\label{Table:StrangeBranchingFractions}
\setlength{\tabcolsep}{0.2pc}
\begin{tabular*}{.67\textwidth}{@{\extracolsep{\fill}}l@{\hspace*{2mm}}l@{\hspace*{2mm}}l} 
\noalign{\smallskip}\hline\noalign{\smallskip}
 $X^-_{us}$                                                & ${\cal{B}}_{\rm{World~Averages}}\, (\%)$ & References                               \\
\noalign{\smallskip}\hline\noalign{\smallskip}
$K^-$ [$\tau$ decay]                                       & $0.690\pm 0.010$                         &                                       \\
\ \ \ \ \ \ ([$K_{\mu 2}$])                                 & ($0.715\pm 0.004$)                       & \cite{Gamiz:2007qs}                  \\
$K^-\pi^0$                                                 & $0.426\pm 0.016$                          & \cite{Aubert:2007jh}                  \\
$\bar{K}^0\pi^-$                                           & $0.835\pm 0.022$ ($S=1.4$)               & \cite{Epifanov:2007rf, Aubert:2008an} \\
$K^-\pi^0\pi^0$                                             & $0.058\pm 0.024$                        & \cite{Yao:2006px}                    \\
$\bar{K}^0\pi^0\pi^-$                                       & $0.360\pm 0.040$                         & \cite{Yao:2006px}                    \\
$K^-\pi^-\pi^+$                                             & $0.290\pm 0.018$ ($S=2.3$)               & \cite{Aubert:2007mh, Inami:2008kt}   \\
$K^-\eta$                                                  & $0.016\pm 0.001$                          & \cite{Inami:2008ar}                   \\
$(\bar{K}3\pi )^-$ (est'd)                                 & $0.074\pm 0.030$                         & \cite{Davier:2005xq}                  \\
$K_1(1270)\rightarrow K^-\omega$                           & $0.067\pm 0.021$                         & \cite{Davier:2005xq}                  \\
$(\bar{K}4\pi )^-$ (est'd)                                 & $0.011\pm 0.007$                         & \cite{Davier:2005xq}                  \\
$K^{*-}\eta$                                                & $0.014\pm 0.001$                       & \cite{Inami:2008ar}                  \\
$K^-\phi$                                                  & $0.0037\pm 0.0003$ ($S=1.3$)           & \cite{Inami:2006vd,Aubert:2007mh}     \\
\noalign{\smallskip}\hline\noalign{\smallskip}
TOTAL                                                     &  $2.8447\pm 0.0688$                       &\\
                                                          & ($2.8697\pm 0.0680$)                      &\\
\noalign{\smallskip}\hline
\end{tabular*}
\end{table}

\subsection{Lepton Universality}
Tests of $\mu-e$ universality can be expressed as:
\begin{eqnarray}
\left( \frac{g_\mu}{g_e} \right)^2 &=& \BRtautomunu \frac{f(m_e^2/m_\tau^2)}{f(m_\mu^2/m_\tau^2)},~\label{emuuniv}
\end{eqnarray}
where $f(x) = 1-8x+8x^3-x^4-12x^2\log{x}$,
assuming that the neutrino masses are negligible.
Eqn.~\ref{br_tautomu_tautoe} yields \gmge = \gmgevalue,
consistent with the previous estimate of $1.000 \pm 0.002$~\cite{Pich:2008ni}.
Tau-muon universality is tested using:
\begin{eqnarray}
\left( \frac{g_\tau}{g_\mu} \right)^2 & = & \frac{\BFtautopinu}{\BFpimutwo} \frac{2m_\pi m^2_\mu\tau_\pi}
                                         {\delta_{\tau^-\to\pi^-\nu/\pi^-\to\mu^-\nu}m^3_\tau\tau_\tau} 
                                         \left( \frac{1-m^2_\mu/m^2_\pi}{1-m^2_\pi/m^2_\tau} \right)^2,~\label{mutauuniv:nonstrange}\\
\left( \frac{g_\tau}{g_\mu} \right)^2 & = & \frac{\BFtautoknu}{\BFKmutwo} \frac{2m_km^2_\mu\tau_K}
                                         {\delta_{\tau^-\to K^-\nu/K^-\to\mu^-\nu}m^3_\tau\tau_\tau} 
                                         \left( \frac{1-m^2_\mu/m^2_K}{1-m^2_K/m^2_\tau} \right)^2, ~\label{mutauuniv:strange}
\end{eqnarray}
where the radiative corrections are
$\delta_{\tau^-\to\pi^-\nu/\pi^-\to\mu^-\nu}  = 1.0016 \pm 0.0014$ and
$\delta_{\tau^-\to K^-\nu/K^-\to\mu^-\nu} = 1.0090 \pm 0.0022$~\cite{Marciano:1993sh}.
Eqn.~\ref{br_tautopi_tautoe} (\ref{br_tautok_tautoe}) yields \gtgm = \gtgmpvalue (\gtgmkvalue) using pions (kaons),
which has a factor of two improvement in precision for kaons than the previous estimate of $0.996 \pm 0.005$ ($0.979 \pm 0.017$)~\cite{Pich:2008ni}.

Multiplying eqn.~\ref{br_tautomu_tautoe} by the world average measured value of $\BFtautoenu$~\cite{Yao:2006px},
and combining with the previous measurements yield a new world average value of $\BFtautomunu = (17.363\pm0.043)\%$.
Assuming $\mu-e$ universality, this result can be used to obtain a predicted value of $\BFtautoenu$.
Assuming $\tau-\mu$ universality, $\BFtautoenu$ can also be obtained from ratio of tau-mu lifetime~\cite{Davier:2005xq}.
Averaging these predictions with the measured value, yields a more precise combined value of $\BFtautoenu^{\rm{univ}} = (17.833\pm0.030)\%$.
Hence, the total hadronic branching fraction is derived as:
$B_{\rm{had}} = 1 - 1.97257 \times \BFtautoenu^{\rm{univ}} = (64.823 \pm 0.059)\%$.
Normalized to the electronic branching fraction, this gives a total hadronic width of $R_{\tau,\rm{had}} = 3.6350 \pm 0.0094$,
assuming lepton universality.

\subsection{{\boldmath $|V_{us}|$} using inclusive strange {\boldmath $\tau$} decays}
The hadronic width of the \mtau can be written as $R_{\tau,\rm{had}} = R_{\tau,\rm{non-strange}} + R_{\tau,\rm{strange}}$.
We can then measure
\begin{eqnarray}
|V_{us}| &=& \sqrt{R_{\tau,\rm{strange}}/\left[\frac{R_{\tau,\rm{non-strange}}}{|V_{ud}|^2} -  \delta R_{\tau,theory}\right]}.
\end{eqnarray}
Here, we use $\Vud = \Vudvalue$~\cite{Eronen:2007qc}, and $\delta R_{\tau,theory} = 0.240 \pm 0.032$~\cite{Gamiz:2006xx} 
obtained with the updated average value of $\ms(2\gev) = 94 \pm 6 \mevcc$~\cite{Jamin:2006tj}.
Normalizing the total strange branching fraction from Table~\ref{Table:StrangeBranchingFractions} to $\BFtautoenu^{\rm{univ}}$,
and using $R_{\tau,\rm{had}}$ from section 2.2 to obtain $R_{\tau,\rm{non-strange}}$,
we measure $\Vus =  0.2159 \pm 0.0030$.
Using the predicted value of $\BR(\tau^- \to K^- \nu_\tau)$ from $K_{\mu 2}$ decays instead of the measured one, we get $\Vus = 0.2169 \pm 0.0029$.

It should be noted that the OPE convergence for this calculation has sizeable instability w.r.t energy scale close to $m_\tau$,
which can be improved using non-spectral weights~\cite{Maltman:2008ib}. 
However, these non-spectral weights also suggest that
\Vus is $\sim 3$ standard deviations lower than the value of $\Vus =  0.2262 \pm 0.0011$
obtained by using the unitarity constraint from the value of $|V_{ud}|$ mentioned above,
as well as from measurements of \Vus from $K_{l2}$ and $K_{l3}$ decays.

\subsection{{\boldmath $|V_{us}|$} using {\boldmath $\BFtautoknu/\BFtautopinu$}}
Encapsulating all non-perturbative effects in the precisely known ratio $f_K/f_\pi = 1.189 \pm 0.007$~\cite{Follana:2007uv}, we measure
\begin{eqnarray}
\frac{\BFtautoknu}{\BFtautopinu} &=& 
\frac{f_K^2 |V_{us}|^2}{f_\pi^2 |V_{ud}|^2} \frac{\left( 1 -  \frac{m_K^2}{m_\tau^2} \right)^2}{\left( 1 -  \frac{m_\pi^2}{m_\tau^2} \right)^2}  
\times \frac{\delta_{\tau^-\to K^-\nut}}{\delta_{\tau^-\to\pi^-\nut}}
~\label{VusBpik}
\end{eqnarray}
to be \BRkpi. In this formulation, the short-distance electroweak contributions cancel.
The long-distance contributions $\delta = \frac{\delta_{\tau^-\to K^-\nut}}{\delta_{\tau^-\to\pi^-\nut}}$ are estimated to be
$\frac{\delta_{\tau^-\to K^-\nu/K^-\to\mu^-\nu}}{\delta_{\tau^-\to\pi^-\nu/\pi^-\to\mu^-\nu}} \times 
\frac{\delta_{K^-\to\mu^-\nu}}{\delta_{\pi^-\to\mu^-\nu}}$, 
where the latter ratio is $0.9930 \pm 0.0035$~\cite{Marciano:2004uf}.
Estimating $\frac{\delta_{\tau^-\to K^-\nu/K^-\to\mu^-\nu}}{\delta_{\tau^-\to\pi^-\nu/\pi^-\to\mu^-\nu}}$ from Ref.~\cite{Marciano:1993sh}, 
gives $\delta = 1.0003 \pm 0.0044$.
Thus, we obtain $\Vus = 0.2254 \pm 0.0023$, which is consistent with the value predicted by unitarity.

\section{SEARCH FOR SECOND CLASS CURRENT}
The conservation of the approximate isospin symmetry implies that 
the hadronic current corresponding to $J^{PG}= 0^{++},0^{--},1^{+-},1^{-+}~(0^{+-},0^{-+},1^{++},1^{--})$ is favored (strongly suppressed),
and known as the First (Second) Class Current, hereafter referred to as FCC (SCC)~\cite{Weinberg:1958ut}.
The decay $\tau^-\to\pi^-\omega\nut$ proceeds dominantly through FCC $(J^{PG}=1^{-+})$ 
mediated by the $\rho$ resonance occurring through a P-wave.
However, it may also proceed through SCC $(J^{PG}=0^{-+},1^{++})$ mediated 
by the $b_1 (1235)$ resonance occurring through S- and D-waves~\cite{Leroy:1977pq}.

These contributions can be studied using the distribution ($F$) of the angle $\theta_{\pi\omega}$
between the normal to the $\omega$ decay plane and the direction of the remaining $\pi^-$ in a $\tau^-\to\pi^-\omega\nut$ decay
measured in the $\omega$ rest frame. While the FCC contribution is $F_{LJ=11}^{FCC}\propto (1-\mbox{cos}^2\theta_{\pi\omega})$,
the SCC contributions are: $F_{LJ=01}^{SCC}\propto 1$, $F_{LJ=10}^{SCC}\propto \mbox{cos}^2\theta_{\pi\omega}$ and
$F_{LJ=21}^{SCC}\propto (1+3\mbox{ cos}^2\theta_{\pi\omega})$, labeled using the orbital ($L$) and total ($J$) angular momenta quantum numbers.

After subtracting background events and applying efficiency corrections, 
a binned fit to the measured $\cos\theta_{\pi\omega}$ distribution is performed 
on signal events containing $\tau^-\to\pi^-\omega\nut$ decays 
obtained from 347\invfb of \epem annihilation data collected by the \babar\ experiment~\cite{Aubert:2008ga} 
using the following fit function:
\begin{equation}
  F(\cos\theta_{\pi\omega})\,
  = N[(1-\epsilon) F_{LJ=11}^{FCC}(\cos\theta_{\pi\omega})+\epsilon F_{LJ=01}^{SCC}(\cos\theta_{\pi\omega})],
\label{eq:onescc}
\end{equation}
where $N$ is a normalization factor and the parameter $\epsilon$ is the fraction 
of $\tau^-\to\pi^-\omega\nut$ decays that proceed through SCC.
In Eq.\ref{eq:onescc}, only $F_{LJ=01}^{SCC}$ is used for the function
describing the SCC contribution since the shape of this
function gives the most conservative (largest) estimate of $\epsilon = (-5.5\pm5.8\rm{(stat)}^{+0.8}_{-5.5}\rm{(syst)})\times10^{-3}$.
No evidence for SCC is observed, and a 90\% confidence level Bayesian upper limit 
for the ratio of SCC to FCC in $\tau^-\to\pi^-\omega\nut$ decays is set at 0.69\%,
which an order of magnitude improvement over the previous best upper limit of 5.4\%~\cite{Edwards:1999fj}.

\section{SUMMARY}
\babar\ 1-prong measurements are consistent with lepton universality within at most $2.5\sigma$,
where imposes strong constraints on new physics models~\cite{Chizhov:1994db, Loinaz:2004qc}.
The branching ratios for the muon and pion channels are as precise as the PDG average,
while the kaon channel is measured with a factor of two improvement in precision.

\Vus from inclusive \mtau decays is $\sim 3 \sigma$ lower than the unitarity prediction. 
Correlations between different measurements are not yet available. 
Also, $(\bar{K}3\pi )^-$ and $(\bar{K}4\pi )^-$ input uses theoretical estimates only. 
Convergence of the integrated OPE series can be improved with non-spectral weights after
measuring the full spectral density function.

\Vus obtained from $\BFtautoknu/\BFtautopinu$ is consistent with unitarity prediction, but individually both
the branching fractions are lower than the universality predictions.

The upper limit on the SCC contribution to $\tau^-\to\pi^-\omega\nut$ decays is improved by an order of magnitude.


\end{document}